\documentclass[11pt]{iopart}
\usepackage{bm}
\usepackage{hyperref}
\usepackage{amssymb}
\usepackage{amsopn}
\usepackage{graphicx}

\renewcommand{\etal}{\emph{et al.}}
\newcommand{\be}{\begin{equation}}
\newcommand{\ee}{\end{equation}}
\newcommand{\fnl}{f_{\mathrm{NL}}}
\newcommand{\Ps}{\mathcal{P}}
\newcommand{\Psw}{\mathcal{P}_{\mathrm{w}}}
\newcommand{\sigmaw}{\sigma_{\mathrm{w}}}
\newcommand{\Mp}{M_{\mathrm{P}}}

\newcommand{\zetawf}{{\zeta_{\mathrm{w}}}}

\renewcommand{\d}{\mathrm{d}}
\newcommand{\vect}[1]{\bm{\mathrm{{#1}}}}
\renewcommand{\e}[1]{\mathrm{e}^{{#1}}}

\renewcommand{\geq}{\geqslant}

\DeclareMathOperator{\AiryAi}{Ai}
\DeclareMathOperator{\AiryBi}{Bi}
\DeclareMathOperator{\skw}{Skew}

\makeatletter
\newcommand\numberwithin[2]{\@addtoreset{#1}{#2}}
\makeatother
\numberwithin{footnote}{section}

\begin{document}
	\title{Non-Gaussianity constrains hybrid inflation}
	\date{\today}
	
	\author{David Mulryne$^{1,2}$, David Seery$^1$ and Daniel Wesley$^{1,3}$ }

	\address{$^1$ Centre for Theoretical Cosmology \\
	Department of Applied Mathematics and Theoretical Physics \\
	Wilberforce Road, Cambridge, CB3 0WA, United Kingdom}
	
	\address{$^2$ Theoretical Physics Group, \\
	Imperial College, London, SW7 2BZ, United Kingdom}

	\address{$^3$ Center for Particle Cosmology\\
	David Rittenhouse Laboratory, University of Pennsylvania \\
	209 South 33rd Street, Philadelphia, PA 19104 USA}	

	\eads{\mailto{D.Mulryne@damtp.cam.ac.uk},
	\mailto{djs61@cam.ac.uk},
	\mailto{dwes@sas.upenn.edu}}

	\submitto{JCAP}

	\pacs{98.80.-k, 98.80.Cq, 11.10.Hi}

	\begin{abstract}
		In hybrid inflationary models, inflation ends
		by a sudden instability associated with 
		a steep ridge in the potential.  
		Here we argue that this feature can generate a large 
		contribution to the curvature perturbation 
		on observable scales. 
		This contribution is almost scale-invariant but highly
		non-Gaussian.	 The degree of
		non-Gaussianity can exceed current observational bounds, 
		unless the
		inflationary scale is extremely low or the hybrid potential 
		contains very large coupling constants.
		Non-linear effects on small scales may quench the non-Gaussian
		signal, and while we find no compelling evidence that this occurs,
		full lattice simulations are required to definitively address
		this issue.
		
    \vspace{3mm}
	\begin{flushleft}
		\textbf{Note added}: We now believe that nonlinear effects will
		invalidate the original computation in this paper essentially
		instantaneously after the short-wavelength modes reach the 
		minimum of their potential.
		This means that the mechanism described in this paper
		will not lead to appreciatable curvature perturbations on long 
		wavelengths, and no useful constraints on hybrid inflation will 
		result.  We have inserted a brief calculation on p2 of this manuscript
		to explain this fact, but have otherwise left the manuscript 
		unchanged.
	\end{flushleft}

	\vspace{3mm}
	\begin{flushleft}
		\textbf{Keywords}:
		Inflation,
		cosmological perturbation theory,
		physics of the early universe,
		quantum field theory in curved spacetime.
	\end{flushleft}
	\end{abstract}
	\maketitle
	
	\newpage
	
\begin{center}
	\textbf{Version 2 of this paper withdraws the claim of large $\fnl$}
\end{center}

After further review, we have come to the conclusion that nonlinear effects
cut off the growth of long-wavelength modes nearly instantaneously after the
short-wavelength modes reach the minimum of their potential.  This means that
even the small difference in timescale explored in Section 5 is not sufficient to guarantee the growth of appreciable long-wavelength fluctuations.  Hence we believe that the mechanism originally described in our paper does not lead to the growth of curvature perturbations on large scales.
Below, we present an argument which illustrates this fact.

We isolate the waterfall field $\chi$ and assume it has a potential
of the form:
\begin{equation}
    V(\chi) = \frac{\lambda}{4} \left( \chi^2 - v^2 \right)
= V_0 - \frac{\mu^2}{2} \chi^2 + \frac{\lambda}{4} \chi^4
\end{equation} 
where $\mu^2 =  \lambda v^2$. (Note $\mu$ has a different meaning in this
note than in the rest of the paper).  The equation of motion
for $\chi$ is 
\begin{equation}\label{e:Xeom}
    \ddot \chi - \nabla^2 \chi - \mu^2 \chi + \lambda \chi^3 = 0,
\end{equation}
and holds pointwise.  We are really interested in the equation
of motion for a related quantity, namely $\chi$ ``smoothed" on 
some length scale $L$.  Let us denote this ``smoothed" version of $\chi(x)$
by $s_L [\chi] (x)$.  There are a few different ways to define this 
smoothed variable, such as convolution with a transfer function or via a
cutoff in Fourier space, but all we really need to proceed with the argument
is that the smoothing is linear, so
\begin{equation}\label{e:PP1}
s_L [c_1 \chi_1 + c_2 \chi_2 ] (x) = 
c_1  s_L [\chi_1 ] (x) +
c_2  s_L [\chi_2 ] (x)
\end{equation}
for all constants $c_1$ and $c_2$, 
and that the smoothing is a projection\footnote{This property holds only approximately for convolutions, but the approximation will be sufficently
accurate for large $L$.}, so
\begin{equation}\label{e:PP2}
s_L [ s_L [\chi] ] = s_L [\chi]
\end{equation}
Since the smoothing process is a projection, we can use it to decompose $\chi$ into a ``smoothed," long-wavelength piece $\chi_L$, and a short-wavelength piece $\chi_S$ as follows
\begin{equation}\label{e:Xd}
    \chi = \chi_L + \chi_S
\end{equation}
where $\chi_L(x) = s_L [\chi](x)$ and 
$\chi_S = \chi - \chi_L$.

We can now study how the short-wavelength dynamics affects the long-wavelength ones.  Applying the smoothing process $s_L$ to the 
equation of motion (\ref{e:Xeom}), one finds
\begin{equation}\label{e:s1}
    \ddot \chi_L - \mu^2 \chi_L + \lambda \chi_L^3 + \lambda \Delta = 0
\end{equation}
where
\begin{equation}\label{e:Ddef}
    \Delta = s_L [ \chi^3 ] - \chi_L^3
\end{equation}
We have dropped the gradient term, since $L$ is much larger than the other scales in the problem.  Equation (\ref{e:s1}) means that the smoothed field $\chi$ has the same equation of motion as the unsmoothed field, up to an additional effective term $\lambda \Delta$.  To estimate this term, we use
the decomposition (\ref{e:Xd}) and the properties
(\ref{e:PP1}) and (\ref{e:PP2}) to find
\begin{equation}
    \Delta = 3 \chi_L^2 s_L[ \chi_S ]  
    + 3 \chi_L s_L[ \chi_S^2 ] + s_L[ \chi_S^3 ]
\end{equation}
The first term must vanish because of the definition of the decomposition
 (\ref{e:Xd}).  The last term may be nonzero, but it seems likely that it
should also vanish because of the $\chi \to -\chi$ symmetry in the problem.
This leaves the middle term.  Once the short-wavelength modes have reached
the minimum of the potential, $\chi_S^2 = v^2$  Hence, once the
short scale-wavelength modes have reached the minimum we have
\begin{equation}
    \Delta = 3v^2 \chi_L 
\end{equation}
Plugging this in to (\ref{e:s1}) results in
\begin{equation}
    \ddot \chi_L + 2\mu^2 \chi_L + \lambda \chi_L^3 = 0
\end{equation}
The short-wavelengthe modes have caused the appearance of an effective mass
term in the equation of motion for $\chi_L$.  This term cancels the tachyonic mass term present in the action, and stabilizes the long-wavelength modes.
This means that by the time
the short-wavelength modes have reached the minimum of the potential, the
long-wavelength modes should cease evolving essentially instantaneously.

It is perhaps interesting that if only quadratic terms were present in the 
action (1), then the equations of motion for $\chi$ would be linear and the 
tachyonic mass term would be preserved in the $\chi_L$ equation of motion.  However, for inflation to end the potential must have a minimum, and the nonlienarities required to introduce the minimum also instantaneously cut off the growth of long-wavelength modes once the short-wavelength modes reach the minimum of the potential.

\begin{center}
\textbf{ The correct calculation}
\end{center}

The $\delta N$ formalism can still be used to calculate the contribution to $\zeta$ from the hybrid 
transition, once we account correctly for the short scales modes. When we smooth the universe on large scales, we 
must remember that although the waterfall field $\chi$ averaged on these scales is extremely small, the kinetic 
energy density of the $\chi$ field averaged on these scales has a contribution from modes of all scales, and 
indeed is dominated by the shortest scale modes present. The time taken for the universe smoothed on large scales  to transition from a flat initial hypersurface to a kinetically dominated comoving final hyperspace is therefore 
not given by equation (46), as we had calculated, but equation (46) with $\chi_*$, the initial field smoothed 
on large scales, replaced by $\sigma_\chi^*$, the RMS value of the initial $\chi$ field, with all modes included.  
This quantity is dominated by the shortest scale modes present in the problem, and can be read off from equation 
(43) using (44) with $k$ given by $k_{\rm short}$.  From the point of view of the large scale modes $\sigma_\chi^*$ 
is almost a constant, but has a small cosmic variance given by $\chi_*$. The induced variance in $N$ and 
therefore $\zeta$, due to this cosmic variance in $\sigma_\chi^*$ is therefore given by the expression
\begin{equation}
\delta N = \zeta = \frac{\partial N }{\partial \sigma_\chi*} \chi_*\,,
\end{equation}
where $N$ is given by equation (46) multiplied by $H$ and with $\chi_*$ replaced by $\sigma_\chi^*$ as discussed above.  Therefore, the curvature perturbation on a particular scale is proportional to the field perturbation averaged on that scale, and is therefore exponentially suppressed on the largest scales.

	\newpage

	\section{Introduction}
	\label{sec:intro}

	Recently, a
	rapid accumulation of data
	has made clear
	that the orthodox scenario of structure
	formation---in which small fluctuations
	oscillate under the influence of gravity
	within a almost-uniform primeval plasma---%
	is in excellent agreement with observation. The initial conditions for
	these fluctuations are determined by the primordial
	curvature perturbation, $\zeta$, 
	usually supposed to have been synthesized during
	an earlier cosmic epoch,
	perhaps during inflation or ekpyrosis.
	Many versions of inflation are characterized by slow, smooth
	evolution and yield very Gaussian fluctuations.
	In other versions evolution is a dramatic process, allowing
	larger non-Gaussianities.

	In this paper, we show that significant power and non-Gaussianity can be
	generated when a set of inflationary trajectories
	fall from a ridge in field space.
	Such a process is a crucial element of the hybrid inflationary 
	scenario, introduced by Linde, in which inflation ends through a sudden
	instability \cite{Linde:1991km,Linde:1993cn,Copeland:1994vg}.
	Quantum fluctuations lead to slight differences in field
	values
	at the top of the ridge, causing different regions of
	the universe to follow one of a narrow bundle of trajectories.
	Trajectories near the edge of the bundle are ejected from the ridge
	early in their evolution, whereas those in the core
	remain on the ar\^{e}te longer.
	The dispersion generated in this way induces a
	variation in expansion history
	(``$\delta N$'') from trajectory
	to trajectory within the bundle,
 	resulting in a nearly scale-invariant but highly non-Gaussian spectrum of
	density perturbations.
	A careful analysis of this process reveals the surprising fact that
	the final dispersion in the bundle depends only weakly
	on the initial dispersion. 
	Hence falling from the ridge can have a large effect on the final 
	curvature perturbation, even if the bundle of trajectories is initially 
	very tightly focused.
	
	Ultimately these contributions to the power spectrum can be traced to the
	conversion of isocurvature modes into adiabatic ones during 
	descent from the ridge. 
	In the early days of the inflationary paradigm, it was assumed that the
	curvature perturbation was primarily determined by	 quantum
	fluctuations in the inflaton field ``freezing out'' as successive
	physical scales left the
	cosmological horizon. Later, when inflationary models
	containing two or more fields were constructed,
	it was understood that the curvature
	perturbation could grow or decay as neighbouring horizon volumes
	diverged along adjacent but inequivalent field-space trajectories.
	The evolution is caused by light isocurvature
	degrees of freedom which play no dynamical role while
	cosmic microwave background (CMB)
	scales are leaving the horizon,
	but later come to influence the expansion history of the universe.
	Although models with isocurvature effects are more complicated, they
	admit a richer phenomenology.
	In models with canonically normalized fields
	the reprocessing of isocurvature
	modes can lead to a significant non-Gaussian component of
	$\zeta$ \cite{Maldacena:2002vr,Lyth:2005fi,Seery:2005gb}.
	In the present case, the isocurvature modes are massive while 
	CMB scales leave the horizon and are suppressed by
	cosmological expansion.
	Nevertheless, falling from the ridge enormously amplifies these
	perturbations and can produce a significant effect.

	Our analysis applies to the original hybrid scenario and many
	related models. The qualitative conclusions may apply during
	a phase of two-field ekpyrosis
	\cite{Koyama:2007mg,Koyama:2007ag,Buchbinder:2007at,
	Koyama:2007if,Lehners:2007wc,Lehners:2008my},
	which also depends on dispersion near a ridge.
	The inflationary production of non-Gaussianity
	in dispersing models
	has been source of interest in recent years.
	Alabidi determined the non-Gaussian yield
	in a hybrid model, beginning from a range of locations
	near but not coincident with the ridge
	\cite{Alabidi:2006hg}.
	Byrnes {\etal} performed a similar analysis, showing that
	the initial conditions could be tuned to produce a large effect
	\cite{Byrnes:2008wi,Byrnes:2008zy}.
	Related studies have appeared in the literature
	\cite{Cogollo:2008bi,Rodriguez:2008hy}.
	
	The generation and evolution of fluctuations in hybrid inflation
	was studied by Randall, Solja\v{c}i\'{c} \& Guth
	\cite{Randall:1995dj}, and later
	Garc\'{\i}a-Bellido, Linde \& Wands
	\cite{GarciaBellido:1996ke,GarciaBellido:1996qt}.
	More detailed calculations were undertaken by
	Copeland, Pascoli \& Rajantie \cite{Copeland:2002ku}.
	The field whose instability ultimately ends inflation is termed
	the \emph{waterfall field} and at early times its potential is designed
	so that it is very heavy, causing its perturbations
	to decay.
	Only fluctuations orthogonal
	to the waterfall make an appreciable contribution
	to the number of e-folds, $N$,
	required to
	reach a subsequent surface of uniform energy density.
	The result is
	a spectrum of density perturbations which can be evaluated by
	ignoring the waterfall field.
	In the simplest scenario, where only one
	canonically normalized field $\phi$ is relevant,
	the single-field formula $\delta N \simeq H \delta \phi / |\dot{\phi}|$
	applies.
	There is an added complication, because the waterfall
	field rolls very slowly at the epoch of ejection.
	Therefore fluctuations in $N$
	may be large on scales leaving
	the horizon at that time,
	and one must verify that there is not an unacceptable synthesis
	of topological defects or primordial black holes
	\cite{GarciaBellido:1996qt}.
	The original hybrid model produced a blue spectrum of perturbations,
	but a larger class of so-called ``hill-top'' models
	\cite{Ovrut:1984qp,Boubekeur:2005zm}
	yield red spectra compatible with CMB constraints
	\cite{Komatsu:2008hk}.

	In this paper, we point out that the traditional calculation of
	curvature perturbations in this model should be augmented
	with a new contribution sourced by the waterfall transition.
	While the trajectories remain on the ridge, fluctuations in $\delta N$
	are well-described by conventional
	perturbation theory applied to the transverse directions, as explained
	above.
	At the epoch of ejection, however,
	fluctuations in the waterfall field become important,
	\emph{no matter how small they are}.
	To see that this can be so, note that
	$\delta N$ measured between the central trajectory (which does not
	leave the ridge) and any other trajectory (which falls into
	a non-inflating minimum) tends to infinity.
	Thus, even exponentially suppressed fluctuations can be
	amplified.
	It follows that a tiny fluctuation
	between neighbouring horizon volumes may ultimately give rise to a
	large density fluctuation.

	The basic mechanism by which large non-Gaussianity is generated can
	be illustrated using
	an inverted simple harmonic oscillator. Suppose we consider
	a particle of unit mass and position $x$, for which the Lagrangian
	\begin{equation}
		L = \frac{1}{2}	 \dot x^2 + \frac{1}{2} \omega^2 x^2
	\end{equation}
	describes an inverted harmonic oscillator of natural frequency $\omega$.
	We wish to compute the time $t$ required for the particle to roll 
	down to some specified final
	position, $x_F$, as a function of its initial position
	$x_0$. We assume that $\omega x_F \gg 1$, $x_0 \ll x_F$,
	and that the particle has
	zero initial velocity.
	At late times $x(t) \sim (x_0/2) e^{\omega t}$, so
	\begin{equation}
		t(x_0) = \frac{1}{\omega} \ln	 
		\frac{2 x_F}{|x_0|}
	\end{equation}
	Next we study the
	statistics of $t$ for an ensemble of particles in which $x_0$
	is Gaussian distributed with variance $\sigma^2$, assuming
	$\omega \sigma \ll 1$.	
	The mean rolling time is
	\be
		\langle t \rangle = 
		\int_{-\infty}^{\infty}
		\frac{\e{-x^2_0 / 2\sigma^2}}{\sqrt{2\pi} \sigma}
		\frac{1}{\omega} \ln	 \frac{2 x_F}{|x_0|} 
		\; \d x_0 = \frac{1}{\omega}
		\left[
		\frac{\gamma+ 3 \ln 2}{2} + \ln \left( \frac{x_F}{\sigma}
		\right) \right]
	\ee
	where $\gamma$ is the Euler--Mascheroni constant.	The average time
	$\langle t \rangle$
	depends only weakly
	on the initial distribution width, $\sigma$, with narrower distributions 
	taking longer to roll off the ar\^{e}te.  More surprising results
	obtain if we compute higher statistics of $t$, such as its
	variance
	\be
		\sigma_{t}^2 = 
		\langle (t	- \langle t \rangle )^2
		\rangle 
		= \frac{\pi^2}{8 \omega^2}
	\ee
	third moment
	\be
		\mu_3 = \langle (t - \langle t \rangle )^3 \rangle
		= \frac{7\zeta (3)}{4 \omega^3}
	\ee
	and fourth moment
	\be
		\mu_4 = \langle (t - \langle t \rangle )^4 \rangle
		=	 \frac{7 \pi^4}{64 \omega^4}
	\ee
	where $\zeta(z)$ is Euler's zeta function and
	$\zeta(3) \sim 1.202$ is Ap\'{e}ry's constant.	 
	Unlike $\langle t \rangle$, these moments are 
	completely independent of the initial variance $\sigma^2$, and 
	depend only on the curvature of the potential.
	Therefore, no matter how small the variance associated with $x_0$,
	the variance in arrival times is the same.
	Moreover, one typically measures departures from Gaussianity by computing
	quantities such as the skew 
	$\mu_3/\sigma_{t}^3$, and excess kurtosis
	$\mu_4/\sigma_{t}^4 - 3$. 
	In our case these are pure numbers, of order unity, which are completely 
	independent of $\omega$. Therefore, our purely Gaussian initial
	distribution always becomes highly non-Gaussian through the process of
	rolling off the hill. 
	
	This toy example can be translated into the inflationary context using a
	simple dictionary.	Roughly speaking, the coordinate $x$ represents the
	waterfall field which mediates the end of hybrid inflation,	 the
	end point $x_F$ represents 
	the location of the reheating minimum, and the time
	$t$ represents the number of e-foldings required to
	reach reheating, at which point inflation ends.	 
	The initial Gaussian distribution of release points models the 
	quantum flucutations in the waterfall field just before the hybrid
	transition.
	Fluctuations in the e-folding history are 
	related to the curvature perturbation, $\zeta$.	Therefore, the
	moments of $t$ 
	represent the moments of $\zeta$.	The fact that they are large
	suggests that large non-Gaussianity can be generated by the
	hybrid waterfall.	Of course,
	the full inflationary story is much more complicated,
	and we will support our conclusions with semi-analytic 
	arguments and
	numerical calculations using realistic forms of the inflationary
	potential. However, the basic mechanism is sufficiently general that its
	essence is
	captured by the inverted harmonic oscillator.
	
	This mechanism generates power on all scales.
	In the traditional picture of hybrid inflation, effects occuring
	near the waterfall transition could only generate significant power on
	short scales.
	For example, Randall {\etal} and Garc\'{\i}a-Bellido {\etal}
	identified
	large fluctuations which collapse to black holes or topological
	defects. These
	are generated on the horizon scale at the epoch of ejection,
	yielding a spike in the power spectrum at small wavelengths. 
	In contrast, a careful analysis of the process
	described above
	indicates that its power spectrum is nearly scale-invariant.
	This is possible because the final variance between expansion
	histories depends very weakly on the initial dispersion of release
	points, as illustrated by the toy model above.
	During inflation, long-wavelength
	fluctuations in the waterfall field are inevitably
	produced, although these are strongly suppressed by cosmic
	expansion.	In the picture we employ, these tiny
	fluctuations are encoded in the width of the bundle of
	field-space trajectories followed by cosmological-scale
	patches of the universe.
	After the hybrid transition,
	the tilt of the power spectrum can be determined by
	comparing the dispersion of the bundle of
	trajectories followed by large-scale patches
	with that of horizon-scale patches.
	Since the final dispersion of 
	trajectories is very weakly dependent on the initial dispersion, the
	final dispersion in each bundle is roughly the same. This is the
	hallmark of a scale-invariant process.
	
	Could this new component in the curvature perturbation become
	observationally detectable?  
	If the growth of perturbations during the waterfall 
	is not curtailed it will be amplified indefinitely,
	yielding a bispectrum in conflict with observation.
	However, any hybrid model automatically contains a cut-off on
	the amplification
	which can be achieved.
	While traversing the ridge, small fluctuations push horizon-scale patches
	to one side or the other, which roll down and reheat into radiation.
	This so-called tachyonic preheating
	\cite{Felder:2000hj,Felder:2001kt,GarciaBellido:2002aj}
	is non-perturbative, and does not admit a description in terms of
	a homogeneous coarse-grained background field with small fluctuations.
	Beyond some limiting time, of order that required for the field to
	reach the minimum, the evolution of large-scale modes becomes dominated
	by radiation produced during the reheating phase
	rather than the inflationary potential. Because the
	ridge in field space no longer drives trajectories to diverge,
	the growth of isocurvature modes from the waterfall ceases.
	
	Ultimately, numerical simulations may be required to determine whether
	a significant effect can be generated before 
	perturbation theory breaks down.
	One might have expected that a very long time would be required before
	isocurvature fluctuations from the waterfall
	could compete with the adiabatic mode.
	In fact, our analysis suggests that the
	field smoothed on large scales typically
	reaches the reheating minimum only a fraction of
	an e-fold after the field smoothed on the horizon scale.
	Whether a given model can generate acceptably Gaussian fluctuations
	depends on a subtle competition between the
	perturbative growth of isocurvature modes and a
	non-perturbative cut-off from tachyonic preheating.
	In our opinion, it is not clear that the outcome can be decided
	on the basis of perturbative calculations.
	Our aim is to show that there is serious interest in obtaining the answer.
	We hope that the
	quantitative intricacies of the competition can eventually be settled
	using lattice simulations.

	The structure of this paper is as follows.
	In \S\ref{sec:background}, we fix our notation by reviewing
	familiar material concerning inflationary perturbations
	and the $\delta N$ formalism.
	In \S\ref{sec:evolution} we give a phenomenological description
	of evolving field fluctuations 
	near a sharp ridge, and use this to derive
	some approximate expressions for the spectrum and bispectrum
	sourced in the curvature perturbation.
	In \S\ref{sec:validity} we discuss the validity of our calculation.
	We work in natural units where $c = \hbar = 1$,
	and the Planck mass associated with Newton's
	gravitational constant is set to unity, $\Mp = (8 \pi G)^{-1/2} = 1$,
	although $\Mp$ will occasionally be restored to clarify the
	relative magnitude of terms.

	\section{The $\delta N$ formalism}
	\label{sec:background}

	Experience has shown that an especially convenient accounting of
	isocurvature modes can be obtained using the ``$\delta N$ formalism,''
	introduced by Starobinsky \cite{Starobinsky:1986fxa}
	and Stewart \& Sasaki \cite{Sasaki:1995aw},
	and extended beyond linear order by
	Lyth \& Rodr\'{\i}guez \cite{Lyth:2005fi}.
	The $\delta N$ formula is an example of the ``separate universe picture,''
	according to which distant Hubble volumes evolve like separate
	unperturbed universes.
	
	The comoving curvature perturbation, $\zeta$, measures fluctuations
	in the expansion history between spatially disjoint
	regions of the universe.
	According to the separate universe picture it can be
	evaluated using the $\delta N$ formula,
	\begin{equation}
		\label{eq:zeta}
		\zeta(t^c, \vect{x}) = N(\rho^c, t^\ast, \vect{x}) - N(\rho^c, t^\ast)
		\equiv \delta N,
	\end{equation}
	where $t^\ast$ labels an initial spatially flat hypersurface,
	and $t^c$ labels a subsequent uniform energy-density hypersurface
	on which the energy takes a prescribed value $\rho^c$.

	Eq.~\eref{eq:zeta} is the lowest term of an expansion in powers of
	the dimensionless gradient
	$k/aH$ but is otherwise non-perturbative.
	Analytic calculations can frequently be simplified by
	constructing a Taylor expansion of $N(\rho^c, t^\ast, \vect{x})$
	around an arbitrary fiducial trajectory.
	When inserted in expectation values, the high order terms
	in this expansion generate contributions
	which grow with volume,
	describing a
	delicately shifting pattern of correlations in the large-volume
	limit.
	In analogy with the growing ultra-violet
	(``loop'') contributions which govern
	correlations in quantum field theory as one samples
	fluctuations of increasing
	energy, these terms have been interpreted as
	``classical loops'' or ``c-loops'' \cite{Zaballa:2006pv,Lyth:2006qz},
	and it has become popular to frame $\delta N$ calculations in this
	language.
	In this paper we do not make use of a loop expansion, but
	apply the non-perturbative definition, Eq.~\eref{eq:zeta}, directly.
	We believe this choice has important advantages.
	First, issues associated with convergence
	are avoided.
	Second, the calculation need not be artificially truncated
	at a low order in the expansion.
	Third, it is unnecessary to introduce an infra-red regulator
	or ``factorization'' scale (without physical significance)
	to make intermediate steps of the calculation finite.
	In quantum field theory the loop expansion is unavoidable owing to
	our ignorance of ultra-violet physics.
	In contrast, in many models
	the infra-red physics associated with Eq.~\eref{eq:zeta}
	is under reasonable control
	and in these cases we believe the loop expansion
	is, at best, an avoidable complication.
	In unfavourable cases it may be rather misleading.
	
	In the following sections, we supply expressions which allow
	$N$ to be calculated in a hybrid model.
	Eq.~\eref{eq:zeta} relates the statistical properties of $N$
	to those of the observable quantity, $\zeta$, by the rules
	\begin{eqnarray}
		\label{eq:zeta-two}
		\langle \zeta \zeta \rangle
		=
		\langle N^2 \rangle - \langle N \rangle^2 , \\
		\label{eq:zeta-three}
		\langle \zeta \zeta \zeta \rangle
		=
		\langle N^3 \rangle - 3\langle N \rangle \langle N^2 \rangle 
		+2 \langle N \rangle^3 .
	\end{eqnarray}
	Similar formulae can be written for higher-order correlation functions.

	Eqs.~\eref{eq:zeta-two}--\eref{eq:zeta-three}
	give the variance and skew of a collection of
	independent spacetime volumes
	smoothed on some characteristic scale, $L$.
	They are not spatially dependent.
	To obtain the real- and Fourier-space correlation functions which can
	be compared with observation, one makes a second use of the
	separate universe picture.
	This implies that any two spatially disconnected volumes will
	be uncorrelated. Therefore,
	\begin{equation}
		\langle \zeta_{\vect{x}} \zeta_{\vect{y}} \rangle =
		\sigma^2(L) L^3 \delta(\vect{x} - \vect{y}) ,
		\label{eq:zeta-two-real}
	\end{equation}
	where we have written $\langle \zeta \zeta \rangle_L = \sigma^2(L)$.
	In Fourier space, the power spectrum $P(k)$ is defined by
	\begin{equation}
		\langle \zeta_{\vect{k}_1} \zeta_{\vect{k}_2} \rangle =
		P(k_1) (2\pi)^3 \delta(\vect{k}_1 + \vect{k}_2) ,
	\end{equation}
	and after making the identification $k \sim L^{-1}$ we can
	conclude that $P(k) = \sigma^2(k)/k^3$.
	The bispectrum, $B(\vect{k}_1, \vect{k}_2, \vect{k}_3)$,
	can be obtained by similar means. It satisfies
	\begin{equation}
		\langle \zeta_{\vect{k}_1} \zeta_{\vect{k}_2} \zeta_{\vect{k}_3} \rangle
		=
		B(\vect{k}_1, \vect{k}_2, \vect{k}_3)
		(2\pi)^3 \delta(\vect{k}_1 + \vect{k}_2 + \vect{k}_3) .
		\label{eq:bispectrum-def}
	\end{equation}
	It is conventional to use momentum conservation
	to make $B$ a function
	of the magnitudes $\{ k_1, k_2, k_3 \}$ alone.
	We write $\langle \zeta \zeta \zeta \rangle_L = \alpha(L)$.
	Then,
	\begin{eqnarray}
		\fl\nonumber
		\langle \zeta_{\vect{x}} \zeta_{\vect{y}} \zeta_{\vect{z}} \rangle
		=
		\frac{\alpha(L)}{3}
		\\ \hspace{-1.8cm} \mbox{} \times \left\{
			L_2^3 L_3^3 \delta(\vect{x} - \vect{y}) \delta(\vect{x} - \vect{z})
			+
			L_1^3 L_3^3 \delta(\vect{y} - \vect{x}) \delta(\vect{y} - \vect{z})
			+
			L_1^3 L_2^3 \delta(\vect{z} - \vect{x}) \delta(\vect{z} - \vect{y})
		\right\} .
		\label{eq:zeta-three-real}
	\end{eqnarray}
	Comparing Eqs.~\eref{eq:bispectrum-def} and~\eref{eq:zeta-three-real},
	it follows that the bispectrum can be written
	\begin{equation}
		B = \frac{\alpha}{3} \frac{\sum_i k_i^3}{\prod_j k_j^3} ,
		\label{eq:local-bispectrum}
	\end{equation}
	which is the `local' form typically generated by superhorizon evolution.
	This is a natural consequence of Eqs.~\eref{eq:zeta-two-real}--%
	\eref{eq:zeta-three-real}, which require that correlations are local
	in real space.
	
	To measure the amplitude of three-point
	correlations it is conventional to use the $\fnl$ parameter, defined by
	\begin{equation}
		B(k_1, k_2, k_3) =
		\frac{6}{5} \fnl \Big\{
			P(k_1) P(k_2) +
			P(k_1) P(k_3) +
			P(k_2) P(k_3)
		\Big\} .
		\label{eq:fnl-def}
	\end{equation}
	Eqs.~\eref{eq:local-bispectrum} and~\eref{eq:fnl-def} imply that
	$\fnl$ satisfies
	\begin{equation}
		\fnl = \frac{5}{18} \frac{\alpha(L)}{\sigma^4(L)} =
		\frac{5}{18}
		\frac{\langle \zeta \zeta \zeta \rangle_L}
		{\langle \zeta \zeta \rangle_L^2} .
	\end{equation}
	This formula can be derived from many alternative constructions
	(see Ref.~\cite{Mulryne:2009kh}).

	\section{Conversion of isocurvature fluctuations at the waterfall}
	\label{sec:evolution}
	
	Consider any region of field space described by
	a collection of light scalars $\phi^\alpha$ and a waterfall
	field $\chi$, evolving according to
	the action
	\begin{equation}
		S = - \frac{1}{2} \int \d^4 x \; \sqrt{-g} \Big\{
				\partial \phi^\alpha \partial \phi_\alpha
				+ (\partial \chi)^2 + 2 V(\phi^\alpha, \chi)
			\Big\} ,
	\end{equation}
	where
	$(\partial\chi)^2 \equiv \partial^a \chi \partial_a \chi$
	and lower-case Latin labels
	$\{ a, b, \ldots \}$ index spacetime dimensions.
	The potential $V(\phi^\alpha, \chi)$ should be
	suitable for realising hybrid inflation but is otherwise arbitrary.
	
	Near the point at which the waterfall field, $\chi$, becomes tachyonic,
	many hybrid potentials can be well-approximated by
	\begin{equation}
		\label{potSimple}
		V = \hat{V}(\varphi, \ldots) - \lambda \varphi \chi^2 + ....
	\end{equation}
	where $\hat{V}$ involves only fields orthogonal to $\chi$.
	The field $\varphi$ is a single direction in field space
	which behaves as an order parameter controlling the onset
	of the waterfall,
	and $\lambda$ is a coupling with the dimensions of mass.
	If the density perturbation is dominated by adiabatic fluctuations
	among these fields, then the spectral tilt, $n_s$,
	depends on $\hat{V}$ alone.
	It is given by
	\begin{equation}
		n_s - 1 = - 6 \epsilon + 2 \eta ,
	\end{equation}
	where the slow-roll parameters are defined by
	\begin{equation}
		\epsilon \equiv
		\frac{\Mp^2}{2} \left( \frac{\hat{V}'}{\hat{V}} \right)^2
		\qquad
		\mbox{and}
		\qquad
		\eta \equiv \Mp^2 \frac{\hat{V}''}{\hat{V}} .
	\end{equation}
	A prime $'$ denotes a derivative with respect to the adiabatic
	direction $\varphi$.
	Linde's original formulation of hybrid inflation
	yielded a blue spectrum \cite{Linde:1991km}, which is now known
	to be incompatible with observation.
	In hybrid models $\epsilon$ is typically negligible,
	so acceptable $\hat{V}$ generally require
	$\eta$ to be slightly negative,
	satisfying $\eta \approx -0.02$. Such models are
	sometimes described as ``hill-top'' potentials
	\cite{Boubekeur:2005zm}.
	Potentials with a negative $\eta$ complicate the higher-order
	terms necessary to guarantee a consistent model, but are conceptually
	no different to the original hybrid proposal. We will use an explicit
	example of such a potential in \S\ref{sec:statistics}.

	\subsection{Analytic approximation}
	\label{sec:analytics}
	
	The dynamics of
	Eq.~\eref{potSimple} can be complicated.
	Our analysis relies on certain simplifying approximations,
	and closely parallels that of Copeland {\etal}~\cite{Copeland:2002ku}.
	We ignore Hubble damping and assume that
	$\dot{\varphi}$ is practically constant.
	These will be acceptable during ejection
	from the ridge, provided that $\varphi$ is rolling 
	slowly and the trajectory is ejected over a timescale
	much shorter than a Hubble time.
	Ultimately, we will
	justify these approximations by checking our results numerically.
	For this purpose we use
	\emph{both}
	the effective potential, Eq.~\eref{potSimple},
	and global completions which describe the descent into a reheating
	minimum, to be discussed below.
	
	It is convenient to work in Fourier space, where
	$\chi$ can be decomposed into a series of coupled oscillators
	$\chi_{\vect{k}}$.
	We may freely choose coordinates so that the effective potential for
	$\chi$ becomes tachyonic at $\varphi = 0$, and set $t=0$ at that time.
	For $t>0$ the evolution of each $\chi_{\vect{k}}$
	is governed by a time-dependent equation of motion,
	\begin{equation}
		\ddot{\chi}_{\vect{k}} + (k^2 - \mu^3 t) \chi_{\vect{k}} = 0 ,
		\label{eq:sigma-eom}
	\end{equation}
	where the linear growth with $t$ is a consequence of our assumption
	that $\dot{\varphi}$ is practically constant.
	The overdot indicates a derivative
	with respect to $t$, and $\mu$ has dimensions of mass,
	\begin{equation}
		\label{eq:mu-def}
		\mu^3 \equiv 2 \lambda \dot{\varphi} \big{|}_{t=0} .
	\end{equation}
	The general solution is
	\begin{equation}
		\label{GenSol}
		\chi_{\vect{k}} (t) =
		A_{\vect{k}} \AiryAi (\mu t - k^2/\mu^2)
		+
		B_{\vect{k}} \AiryBi (\mu t - k^2/\mu^2) ,
	\end{equation}
	where
	$A_{\vect{k}}$ and $B_{\vect{k}}$ are arbitrary coefficients,
	and $\AiryAi(x)$ and $\AiryBi(x)$ are Airy functions of the first and
	second kind, respectively. For $x<0$ these functions are
	oscillatory, whereas for $x>0$ the first Airy function
	decays and $\AiryBi(x)$ grows exponentially.
	This growth corresponds to an exponential rise in occupation number
	of the low-lying modes of $\chi$. It describes a rapid reordering of
	the spatial field configuration, during which gradients may become
	large. We will return to this issue in \S\ref{sec:validity}.
	
	For a mode of wavenumber $k$, exponential growth begins at the time
	$t_k \equiv k^2/\mu^3$. It follows that the phase transition proceeds
	by spinodal decomposition, beginning with the zero mode
	and extending to higher $k$-modes at successively later times.
	The coefficients
	$A_{\vect{k}}$ and $B_{\vect{k}}$
	can be expressed
	in terms of $\chi_{\vect{k}}(t_k)$ and $\dot{\chi}_{\vect{k}}(t_k)$, giving
	\begin{eqnarray}
		\label{constants}
		A_{\vect{k}} =
		\frac{3^{2/3}}{2} \Gamma(2/3) \chi_{\vect{k}}(t_k)
		- \frac{3^{1/3}}{2 \mu} \Gamma(1/3) \dot{\chi}_{\vect{k}}(t_k) \\
		B_{\vect{k}} =
		\frac{3^{1/6}}{2} \Gamma(2/3) \chi_{\vect{k}}(t_k)
		+ \frac{3^{-1/6}}{2 \mu} \Gamma(1/3) \dot{\chi}_{\vect{k}}(t_k) .
	\end{eqnarray}

	Our interest lies in a description of the process by which
	spatially disjoint spacetime regions
	depart from the ar\^{e}te.
	It is at this point that our analysis departs from
	that of Copeland {\etal}, whose interest lay with the smallest
	possible scales.
	Consider an ensemble of spacetime volumes passing over the
	waterfall, smoothed on a lengthscale
	of order $k^{-1}$.
	The mutual scatter in field value between the members of this ensemble
	can be computed, for which it is a good approximation to
	include only the growing mode. We find
	\begin{equation}
		\sigma^2_\chi(k,t)
		=
		\frac{1}{2\pi^2}
		\int^{k}_0 
		P_B({k'})
		\AiryBi [\mu (t - t_{k'}) ]^2
		{k'}^2 \; \d {k'} ,
		\label{eq:chi-scatter}
	\end{equation}
	where we have
	filtered out modes of wavelength shorter than $1/k$, and
	introduced the power spectrum of $B_k$,
	\begin{equation}
		\langle B_{\vect{k}}^\ast B_{\vect{k}} \rangle
		=
		P_B(k) (2\pi)^3 \delta({\bf k}- {\bf k'}).
	\end{equation}
	Eq.~\eref{eq:chi-scatter}
	is valid when the relevant $k$-modes are already rolling,
	and hence applies provided $k \lesssim \sqrt{\mu^3 t}$.
	We have set an implicit infrared cutoff scale to zero.
	It transpires that $P_B(k)$ is a very blue function of $k$,
	so this is typically an excellent approximation. 

	Determining $P_B(k)$
	requires information about the quantum fluctuations in $\chi$.%
		\footnote{In their analysis,
		Copeland {\etal} noted that modes are light at the
		tachyonic transition and assumed that each mode was in a massless
		Bunch--Davis vacuum state at time $t_k$. In this state,
		the correlation functions satisfy
		\begin{eqnarray}
			\nonumber
			\langle \chi^*_{\vect{k}}\chi_{\vect{k}'}\rangle
			=
			\frac{1}{2k}
			(2 \pi)^3 \delta(\vect{k} - \vect{k}') \\
			\langle \dot{\chi}^*_{\vect{k}} \dot{\chi}_{\vect{k}'} \rangle
			=
			\frac{k}{2}
			(2\pi)^3 \delta(\vect{k}- \vect{k}') . 
		\end{eqnarray}
		This would be a good choice whenever the effective mass $M$ of
		$\chi$ varies adiabatically, or whenever $\dot M / M^2 \ll 1$.
		Very close to the tachyonic transition, $M^2 \sim t$ and evolution
		will not be adiabatic.}
	Immediately prior to the tachyonic transition,
	$\chi$ carries a
	power spectrum imprinted by the preceeding epoch, during which
	its mass was large. It is a reasonable
	approximation to determine the fluctuations synthesized
	during this era by setting each $\vect{k}$-mode
	in a Bunch--Davies vacuum state, with
	large constant mass $M$.
	Well inside the horizon, each mode is normalized
	to the corresponding Minkowksi space oscillator.
	Once $k/a \lesssim M$ the mass term will dominate its evolution, and each
	mode decays as $\chi_{\vect{k}} \propto a^{-3/2}$ owing to
	Hubble damping. For super-horizon modes,
	where $k/a \ll H$, this leads to%
		\footnote{See~\ref{appendix:massive} for details.}
	\begin{eqnarray}
		\label{ic2}
		\nonumber
		\langle \chi^*_{\vect{k}}\chi_{\vect{k}'} \rangle
		=
		\frac{1}{2M}
		\left( \frac{H}{k_\star} \right)^3
		(2\pi)^3 \delta(\vect{k}- \vect{k}') \\
		\langle \dot{\chi}^*_{\vect{k}} \dot{\chi}_{\vect{k}'} \rangle
		=\frac{M}{2}
		\left( \frac{H}{k_\star} \right)^3
		(2\pi)^3\delta(\vect{k}- \vect{k}') ,
	\end{eqnarray}
	where $k_\star$ is the comoving wavenumber of the mode leaving the
	horizon at the time of evaluation.
	These correlators scale like $a^{-3}$, because
	the wavenumber $k_\star$ scales like $a$ as inflation proceeds.
	Employing~\eref{constants} and~\eref{ic2} yields%
		\footnote{We have evaluated the correlators
		$\langle \chi_{\vect{k}}^\ast(t_k) \chi_{\vect{k}'}(t_k) \rangle$
		and
		$\langle \dot{\chi}_{\vect{k}}^\ast(t_k) \dot{\chi}_{\vect{k}'}(t_k)
		\rangle$
		at a common time, $t=0$, rather than $t_k$.
		The error we commit in this approximation is negligible,
		since the e-foldings of expansion between $t=0$ and $t=t_k$
		is of order $(k/k_\ast)^2 (H/\mu)^3 \ll 1$.}
	\begin{equation}
		 \label{eq:PB}
		P_B(k) =	
		\frac{H^3}{8 M k_\star^3}
		\left[
			{3^{1/3} \Gamma(2/3)^2}
			+
			{3^{-1/3} \Gamma(1/3)^2} \left( \frac{M}{\mu} \right)^2
		\right]
		=
		\frac{C H^3}{M k_\star^3} ,
	\end{equation}
	where $C$ aggregates the numerical constants. It is
	at least of order unity, and can be much larger if $\mu \ll M$.  
	
	For large $x$,
	the growing Airy function $\AiryBi(x)$ takes the asymptotic form
	\begin{equation}
		\label{asyBi}
		\AiryBi(x)
		\simeq \frac{1}{\sqrt{\pi}}
		x^{-1/4} \exp\bigg( \frac{2}{3} x^{3/2} \bigg) .
	\end{equation}
	At late times, this implies that
	the scatter among field values is well-approximated by
	\begin{equation}
		\label{varianceInt}
		\sigma^2_\chi(k,t) \approx
		\frac{C H^3}{2\pi^3 M k_\star^3}
		\int^k_0
		\frac{1}{\sqrt{\mu (t-t_{k'})}}
		\exp \left(\frac{4}{3} \Big[\mu (t-t_{k'}) \Big]^{3/2} \right )
		{k'}^{2} \, \d {k'} .
	\end{equation}
	When $t \gg t_k$ the $k$-dependence becomes trivial: all modes
	behave collectively, evolving coherently with the background field.
	Moreover, the $k$-integral can be performed
	at once. In this limit, we find
	\begin{equation}
		\label{varianceChi}
		\sigma^2_\chi(k,t) = 
		\frac{C H^3 k^3}{6 \pi^3 M k_\star^3}
		\frac{1}{\sqrt{\mu t }} \exp \left(\frac{4}{3} [\mu t]^{3/2} \right ) .
	\end{equation}
	Eq.~\eref{varianceChi} implies that the scatter
	within an ensemble of
	spacetime volumes %traversing the waterfall
	grows according to the
	equation of motion for the background field in a single volume,
	\begin{equation}
		\label{asyWaterSol}
		\chi = \frac{\chi_\ast}{(\mu t)^{1/4}}
		\exp\left(
			\frac{2}{3} [ \mu t ]^{3/2}
		\right) .
	\end{equation}
	This justifies our application of the separate universe
	picture, but the preceding Fourier space analysis is required
	to determine
	the effective release point, $\chi_\star$,
	as a function of scale.
	This can be obtained from inspection of Eq.~\eref{varianceChi}.

	\subsection{The statistics of $N$}
	\label{sec:statistics}
	
	Once the waterfall potential has become tachyonic, $\chi$ rolls
	down %one side of
	the ridge. Its velocity increases until
	the universe becomes dominated by the kinetic energy $\dot{\chi}^2/2$.
	When this phase of ``kinetic domination'' is achieved,
	comoving hypersurfaces are practically determined by hypersurfaces
	of constant kinetic energy, and $\zeta$ ceases to evolve.
	According to Eq.~\eref{asyWaterSol},
	the distribution of initial values, $\chi_\ast$, will lead to a spread
	in arrival times at the kinetically dominated epoch.
	We write the perturbation in expansion history associated with this
	spread as $\zetawf$. It is this
	fluctuation which we propose must be included when calculating the
	properties of perturbations generated in a typical hybrid model.
	The following
	perturbative analysis will be trustworthy only if kinetic
	domination is reached. In \S\ref{sec:validity} we discuss the
	possibility that a non-perturbative effect, \emph{tachyonic preheating},
	can quench the dispersion of trajectories before kinetic domination
	is attained.
	
	The transit time to kinetic domination,
	$t(\chi_\ast)$, can be estimated by requiring
	$\dot{\chi} \sim H$. Taking $H$ to be constant,
	Eq.~\eref{asyWaterSol} implies that $t(\chi_\ast)$ must solve
	\begin{equation}
		\fl
		\ln \frac{H}{\chi_\ast} =
		-\frac{1}{4} \ln \Big[\mu t(\chi_\ast)\Big] +
		\ln \left(
			\frac{\mu}{\Mp} \sqrt{\mu t(\chi_\ast)}
			- \frac{1}{4t(\chi_\ast) \Mp}
		\right)
		+
		\frac{2}{3} \Big[ \mu t(\chi_\ast) \Big]^{3/2} ,
	\end{equation}
	in which the Planck mass, $\Mp$, has temporarily been restored
	to exhibit the relative magnitude of each term.
	The asymptotic regime of late times corresponds to
	$\mu t \gg 1$, and is approximately reached when
	$N \sim H/\mu$ e-folds have elapsed since the
	waterfall transition.
	In this region $\ln (\mu t)$ can be neglected in comparison with
	$(\mu t)^{3/2}$.
	Under the same assumptions, $(t \Mp)^{-1}$ can be neglected in
	the middle logarithm, after which
	the remaining term is of order $\ln \mu \Mp^{-1}$.
	This may be
	large if $\mu$ is much smaller than the Planck scale.
	On the left-hand side, however,
	$\ln H \chi_\ast^{-1}$ is typically of order
	$1.5 \ln (k_\ast/k) + 0.5 \ln M H^{-1} \sim 100$.
	To avoid fatal problems with overproduction of black holes and
	topological defects, discussed
	by Garc\'{\i}a-Bellido {\etal} \cite{GarciaBellido:1996qt},
	the entire waterfall phase should complete within an e-fold.
	Since the scale for roll-down is set by $\mu$, this requires
	$\mu \gtrsim H$. Hence $\mu$ cannot be 
	arbitrarily small,
	and  $|\ln (\mu \Mp^{-1})|$ is likely to be much smaller than 100,
	and can be discarded in comparison with $\ln H \chi_\ast^{-1}$.
	The remaining terms imply
	that the transit time
	is given to a fair approximation by
	\begin{equation}
		t(\chi_\ast) = \frac{1}{\mu} \left(
			\frac{3}{2} \ln \frac{H}{\chi_\ast}
		\right)^{2/3} .
		\label{eq:transit-time}
	\end{equation}
	Note that
	the typical smallness of $\chi_\ast$, which at first sight might
	cause one to imagine that any isocurvature fluctuation is tiny,
	has played an important role in ascertaining
	the rough validity of Eq.~\eref{eq:transit-time}.

	In virtue of our assumption that $H$ is practically constant,
	Eq.~\eref{eq:transit-time} can be converted to an estimate of the
	transit time in e-folds, $N(\chi_\ast)$.
	The moments of $N$ satisfy
	\begin{eqnarray}
		\label{eq:n-one}
		\langle N \rangle
		=
		\frac{H}{\mu}
		\int^{\infty}_0
		P(\chi_*, \sigma_\chi)
		\left(\frac{3}{2} \ln \frac{H}{\chi_*} \right )^{2/3}
		\d \chi_\ast ,
		\\
		\label{eq:n-two}
		\langle N^2 \rangle
		=
		\left(\frac{H}{\mu}\right)^{2}
		\int^{\infty}_0
		P(\chi_*, \sigma_\chi)
		\left( \frac{3}{2} \ln \frac{H}{\chi_*} \right )^{4/3}
		\d \chi_\ast ,
		\\
		\label{eq:n-three}
		\langle N^3 \rangle
		=
		\left(\frac{H}{\mu}\right)^{3}
		\int^{\infty}_0
		P(\chi_*, \sigma_\chi)
		 \left (\frac{3}{2} \ln \frac{H}{\chi_*} \right )^{2}
		\d \chi_\ast ,
	\end{eqnarray}
	where $P(\chi_\ast, \sigma_\chi)$ is the distribution of initial
	values, $\chi_\ast$.
	It is determined by the quantum mechanics of a massive scalar
	field evolving under Hubble damping.
	Therefore, we expect $P$ to be %almost exactly
	close to Gaussian,
	characterized by its variance but with negligible higher moments.
	
\begin{figure}
		\center{\includegraphics[width = 10cm]{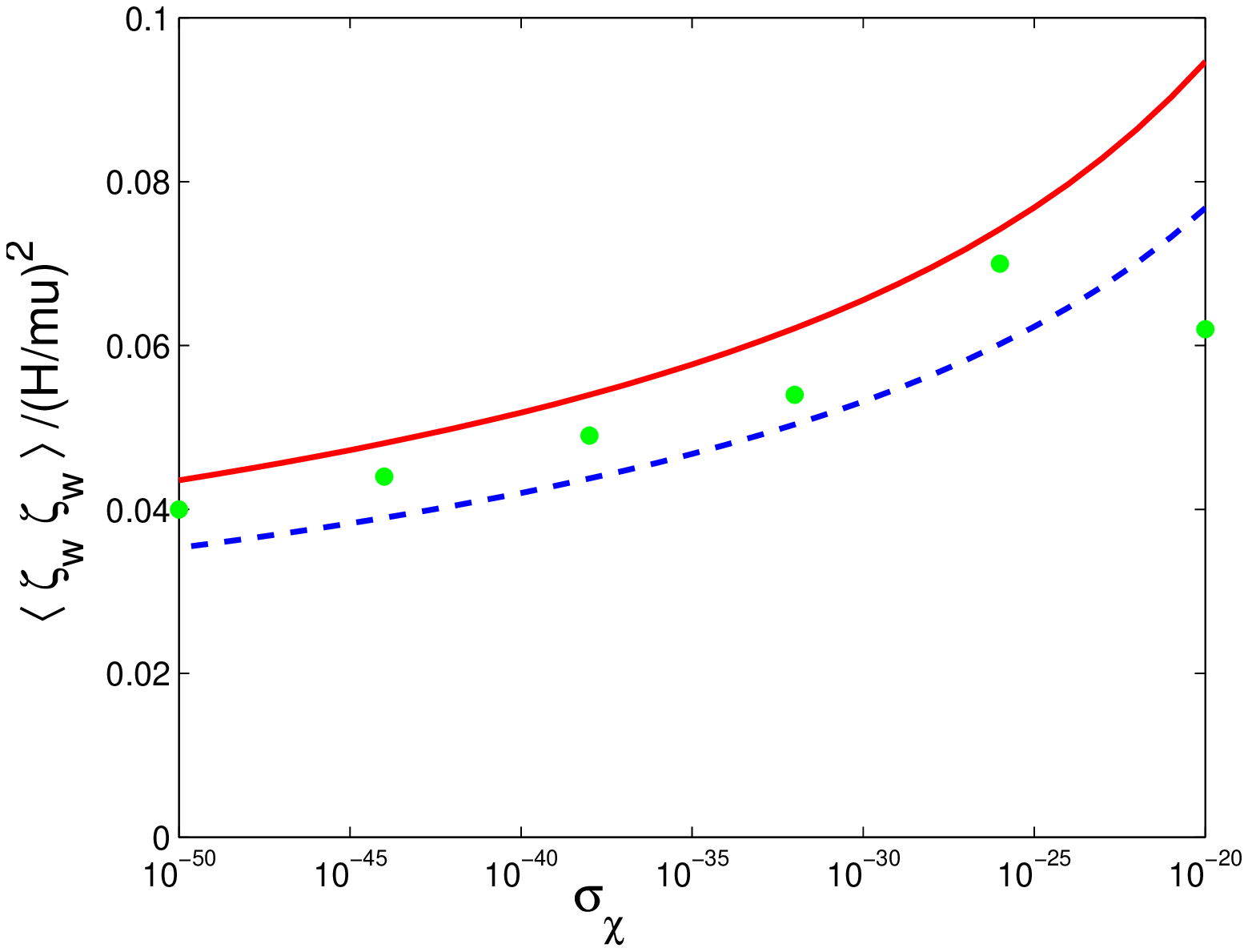}}
		\caption{\label{fig1}
			$\sigmaw^2/(H/\mu)^2$  plotted as a function of the RMS 
			initial fluctuation
			$\sigma_\chi$ in $\chi$.	 The green dots are	 
			computed numerically for a Gaussian ensemble of trajectories, 
			using the potential \eref{e:numV} and parameter
			values \eref{e:numVparams}.	The lines are semi-analytic
			estimates, obtained by numerically
			integrating (\ref{eq:n-one}-\ref{eq:n-three}).  The solid
			red line uses a Gaussian distribution of initial conditions of
			variance $\sigma_\chi^2$,
			while the dashed blue line employs a uniform distribution of
			varaince $\sigma_\chi^2$.  The final variance depends weakly on the
			form of the initial distribution, and varies extremely 
			slowly over many decades of the initial 
			variance.
			}			
	\end{figure}

\begin{figure}
		\center{\includegraphics[width = 10cm]{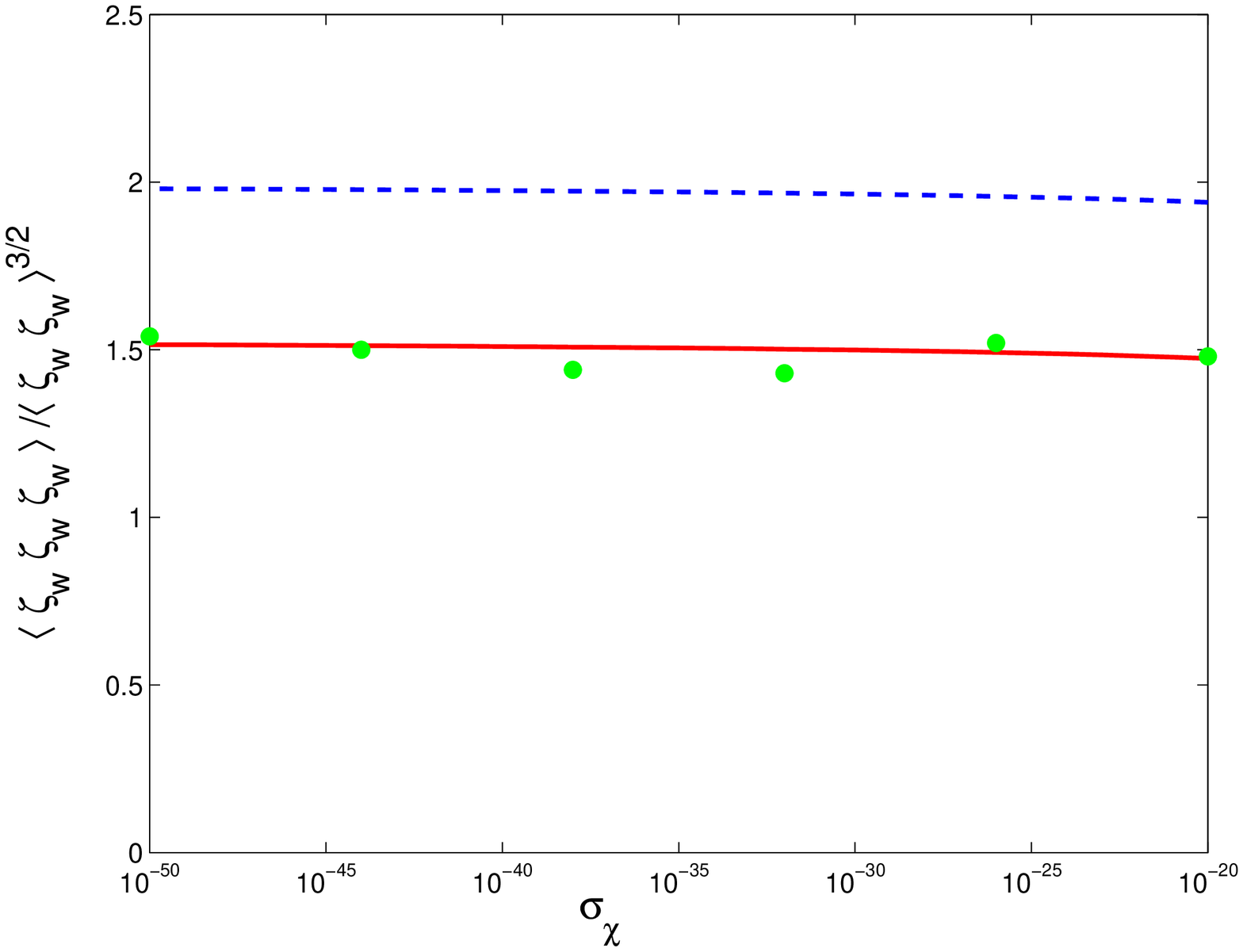}}
		\caption{\label{fig2}
			The skew of $\zetawf$  plotted as a function of the 
			RMS initial fluctuation $\sigma_\chi$ in $\chi$.
			The green dots are  
			computed numerically for a Gaussian ensemble of trajectories 
			using the potential \eref{e:numV} and parameter
			values \eref{e:numVparams}.	The lines are semi-analytic
			estimates, obtained by numerically
			integrating (\ref{eq:n-one}-\ref{eq:n-three}).  The solid
			red line uses a Gaussian distribution of initial conditions of
			variance $\sigma_\chi^2$,
			while the dashed blue line employs a uniform distribution of
			variance $\sigma_\chi^2$.  The final skew depends weakly on 
			the form of the initial distribution, and varies very
			little over many decades of variation in the initial 
			variance.
			}
	\end{figure}

	In analogy with the inverted simple harmonic oscillator
	considered in \S\ref{sec:intro},
	these expressions lead to the surprising result that
	the variance and
	skew of $\zetawf$
	depend only weakly
	on the initial variance,
	$\sigma_\chi^2$. Furthermore,
	even if the initial distribution $P(\chi_*, \sigma_\chi)$,
	is exactly Gaussian, a skew of order unity is generated.
	This behaviour persists over an exponentially large range
	of $\sigma_\chi^2$.
	We have evaluated
	Eqs.~\eref{eq:n-one}--\eref{eq:n-three} numerically, and
	used Eqs.~\eref{eq:zeta-two}--\eref{eq:zeta-three}
	to determine statistics of the observational quantity, $\zetawf$.
	Our results are depicted in Fig.~\ref{fig1}.
	To a good approximation, it is clear that
	$\langle \zetawf \zetawf \rangle = \sigmaw^2 \sim \gamma (H/\mu)^2$,
	where $\gamma$ varies slowly within 
	the range $0.01<\gamma<0.1$ 
	over an exponentially large range of $\sigma_\chi^2$.
	Similar behaviour occurs for the skewness, % of $N$,
	defined by
	\begin{equation}
		\skw(\zetawf) =
		\frac{\langle \zetawf \zetawf \zetawf \rangle}
		{\langle \zetawf \zetawf \rangle^{3/2}}
	\end{equation}
	and displayed in Fig.~\ref{fig2}.
	Over a similar exponentially large range of $\sigma_\chi$
	it is clear that $\skw(\zetawf)$ is practically constant,
	taking values close to $1.5$ for an Gaussian initial distribution of 
	$\chi_*$.

	We have verified these results using numerical Monte Carlo simulations,
	which give good agreement with the semi-analytic results given above.
	We have carried out simulations
	for a simple example of a potential which 
	agrees with Eq.~\eref{potSimple} in the  
	transition region, but which is bounded from below and describes
	the descent into a stable reheating minimum at zero energy.
	The potential we use is 
	\begin{eqnarray}
		\label{e:numV}
		V = V_0 
		\left[
			\left(
				1 + \frac{\eta \phi^2}{4 \Mp^2} 
			\right)
			-
			\frac{\lambda\left(\phi^2-\phi^2_0\right)\chi^2}{4V_0\phi_0}
		\right]^2 \\
		\label{e:numVx}
		\sim V_0 + \frac{\eta	 V_0}{2 \Mp^2} \varphi^2
		- \lambda \varphi \chi^2 + O(\varphi^3,\dots)
		\label{eq:complete-potential}
	\end{eqnarray}
	where $\varphi = \phi-\phi_0$.
	Near the hybrid transition, Eq.~\eref{e:numVx} indicates that 
	\eref{e:numV} has the approximate form of Eq.~\eref{potSimple}.
	In using Eq.~\eref{eq:complete-potential}
	we are free to end the simulation at
	any time after kinetic domination is reached.
	For convenience we 
	choose the time at which the field is close to minimum for the first time. 
	This occurs just after kinetic domination.	 
	We evaluate $N$ as a function of the initial values
	$\phi_\ast$ and $\chi_\ast$ by 
	solving the full cosmological field equations.
	Initially $\phi$ is fixed
	so that the system is on the cusp of the hybrid instability.
	For the full potential, Eq.~\eref{e:numV},
	this corresponds to choosing $\phi_\ast = \phi_0$.
	The initial value, $\chi_\ast$,
	is drawn from a Gaussian distribution with variance 
	$\sigma_\chi^2$.	
	This process is repeated $\sim 10^4$ times, for which
	$N$ is evaluated at a fixed value of $H$.
	This procedure accurately provides
	values for $\langle N \rangle$, $\langle N^2 \rangle$ and 
	$\langle N^3 \rangle$ evaluated on a comoving hypersurface. 
	Eqs.~\eref{eq:zeta-two}--\eref{eq:zeta-three} 
	are employed to determine the statistics of $\zeta$. 
	Physically, this procedure is equivalent to allowing 
	an ensemble of trajectories, with initial conditions characterized by
	a distribution of
	values $\chi=\chi_\ast$, to evolve under the influence
	of the hybrid potential.
	No approximations are made during the dynamical evolution.
	We include the effect of cosmological expansion on the evolution of
	$\phi$ and $\chi$, and
	retain all contributions to the Hubble rate.

	We have performed this proceedure for a wide range of parameter values.
	The results are always in
	good agreement with our semi-analytic formulae. We
	find that the variance $\sigmaw^2$ depends only weakly
	on the initial variance, $\sigma_\chi^2$.
	More importantly, the skew 
	is of order unity, irrespective of the value of $\sigma^2_\chi$.
	Our conclusion that $\zetawf$ carries
	a strongly non-Gaussian distribution appears
	to be 
	robust.
	In Figs.~\ref{fig1}--\ref{fig2}, we 
	present results obtained using the full potential 
	\eref{e:numV} and the specific choices
	\be
		\label{e:numVparams}
			\eta = -0.02, \quad
			\lambda = 10^{-5}, \quad
			V_0 = 10^{-11}, \quad
			\phi_0 = 1.
	\ee
	In this model
	$H/\mu \simeq 0.02$.

	One may have harboured concerns
	the approximation of Eq.~\eref{eq:sigma-eom},
	in which $\dot{\varphi}$ is taken to be constant,
	could be a good approximation near the top of the ar\^{e}te,
	but a poor description of the waterfall evolution near kinetic
	domination, where the dominant contribution to $\zetawf$ is
	synthesized. The good agreement shown in
	Figs.~\ref{fig1}--\ref{fig2} between our exact Monte Carlo
	simulations and the semi-analytic formulae shows that
	Eq.~\eref{eq:sigma-eom} provides a good approximation throughout
	the evolution of the waterfall field.
	
	\subsection{Power spectrum and $\fnl$}
	\label{sec:ps-and-fnl}

	According to the argument of \S\S\ref{sec:analytics}--\ref{sec:statistics},
	we have identified a new source of curvature perturbations in hybrid-like
	scenarios. However, we cannot yet conclude that this
	perturbation is present in typical hybrid models because the foregoing
	analysis depends on the validity of the separate universe picture.
	Our numerical experiments show that the dominant part of
	$\zetawf$ is generated close to the era of kinetic domination.
	In the Introduction (\S\ref{sec:intro}), we discussed the role of
	tachyonic preheating in quenching the dispersion of trajectories.
	This is associated with a failure of the separate universe picture.
	If tachyonic preheating completes before kinetic domination is
	attained on CMB scales, then the large-scale waterfall field
	acquires negligible dispersion.
	In this case the evolution of
	perturbations follows the traditional picture
	of Refs.~\cite{Randall:1995dj,GarciaBellido:1996qt}.%
		\footnote{A significant non-Gaussian perturbation is always generated
		on the horizon scale, which may or may not be competitive
		with the perturbations studied by
		Randall {\etal}
		\cite{Randall:1995dj} and Garc\'{\i}a-Bellido {\etal}
		\cite{GarciaBellido:1996qt}.
		We defer consideration of this effect until \S\ref{sec:conclusions}.}
	We study this possibility in detail in \S\ref{sec:validity}.
	
	In the remainder of \S\ref{sec:evolution} we assume that
	the dynamics of some specific model are such that kinetic domination
	occurs before the onset of gradient instabilities associated with
	tachyonic preheating.
	In any such model $\zetawf$ is uncorrelated with,
	and adds incoherently to,
	the curvature perturbation synthesized during the previous
	inflationary epoch.
	This is similar to the scenario studied by
	Boubekeur and Lyth \cite{Boubekeur:2005fj}, who considered
	the effect of an uncorrelated
	$\chi^2$ perturbation which added incoherently
	to a scale-invariant inflationary contribution.	

	Does $\zetawf$ unacceptably contaminate the primordial density
	perturbation in the hybrid scenario? The answer depends on the spectrum
	of $\zetawf$, and whether it makes a significant contribution on CMB
	scales.
	The spectrum of $\zetawf$ is determined by 
	the dependence of $\sigmaw^2$ on the moments
	which characterize
	the initial distribution of $\chi_\ast$.
	The continuum power spectrum associated with $\zetawf$ takes the form
	\begin{equation}
		\label{PowerZeta1}
		\Psw(k)
		=
		\frac{\d \sigmaw^2}{\d \ln k}
		=
		\frac{\partial \sigmaw^2}{\partial \sigma_\chi^2}
		\frac{\d \sigma_\chi^2}{\d \ln k}
		+
		\sum_{m \geq 3}
			\frac{\partial \sigmaw^2}{\partial \mu_m}
			\frac{\d \mu_m}{\d \ln k} ,
	\end{equation}
	where $\mu_m$ is the $m$th moment of $\chi_\ast$,
	$\sigma_\chi^2$ is the initial variance of $\chi$,
	and $\Ps$ denotes the dimensionless power spectrum,
	defined for any fluctuation by the rule $\Ps(k) = k^3 P(k) / 2\pi^2$.
	Comparison of the Gaussian and uniform distributions in
	Fig.~\ref{fig1} (which have very different higher moments)
	gives reasonable evidence that
	$\d \sigmaw^2 / \d \sigma_\chi^2$ is roughly independent
	of the $\mu_m$.
	There is no reason to believe that
	the $\d \sigmaw^2 / \d \mu_m$
	are small in comparison,
	but if the initial distribution of $\chi$
	is close to Gaussian then 
	these moments can be neglected.
	As we argued above it seems reasonable to expect that
	$P(\chi_\ast)$ is approximately Gaussian, because it is determined
	by the fluctuations of very massive oscillators evolving under
	Hubble damping. Under these assumptions the power spectrum of
	$\zetawf$ is approximately
	\begin{equation}
		\Psw(k)
		\simeq
		\frac{\d \sigmaw^2}{\d \sigma^2_\chi}
		\Ps_{\chi}(k) .
		\label{eq:zeta-power}
	\end{equation}
	
	In view of the steep, blue spectrum associated with
	$\chi_\ast$ one might worry that Eq.~\eref{eq:zeta-power}
	implies $\Psw$ is also very blue.
	This turns out not to be the case.
	A slowly-varying function such as $\sigmaw^2(\sigma_\chi^2)$ can be
	well-approximated by a series in powers of $\ln (H/\sigma_\chi)$, of
	the form
	\begin{equation}
		\label{sigmaZeta}
		\sigmaw^2
		=
			A_0 \left[ \ln \left(\frac{H}{{\sigma_\chi}}\right) \right]^{-p}
		+ A_1 \left[ \ln \left(\frac{H}{{\sigma_\chi}}\right) \right]^{-p-1}
		+ \cdots +
		\Or\left( \frac{\sigma_\chi}{H} \right) .
	\end{equation}
	The parameter $p$ must be strictly positive, because
	$\sigmaw^2$ must be an increasing function of $\sigma_\chi^2$.
	Subsequent terms in
	the series are subleading.
	One can show analytically that $\sigmaw^2$ takes precisely this form
	when the distribution for $\chi_\ast$ is uniform,
	with leading term equal to 
	$(2/3)^{2/3} (H/\mu)^2 \left[ \ln (H/\sigma_\chi) \right]^{-2/3}$.
	Focusing on the leading term, one obtains a power spectrum 
	\begin{equation}
		\Psw(k)
		=
		\frac{\d \ln \sigma_\chi^2}{\d \ln k}
		\frac{(p A_0)}{2} \left[
			\ln \left (\frac{{H}}{\sigma_\chi} \right)
		\right]^{-p-1} + \cdots .
		\label{eq:zetawf-ps}
	\end{equation}
	Eq.~\eref{eq:zetawf-ps} shows that
	$\Psw(k)$ is proportional to the tilt of the  power spectrum of
	$\Ps_\chi(k)$,
	rather then the input spectrum itself. 
	Even if $\chi$ has an extremely steep power spectrum, provided
	its running is small, the power spectrum of
	$\zetawf$ is approximately scale invariant.
	Had we retained subleading terms from \eref{sigmaZeta}, 
	proportional to higher powers
	of $[ \ln H/\sigma_\chi ]^{-1}$, the same conclusion would have been
	obtained. Such terms may modify the normalization of $\Psw$
	but do not change its shape.
	
	Our next step is to
	estimate $\fnl$. Our semi-analytic discussion and the
	numerical evidence shows that the skew of $\zetawf$ is
	of order unity over a
	very wide range of $\sigma_\chi$, and 
	is essentially independent of $k$. It follows that
	\begin{equation}
		\label{eq:zetawfskew}
		\langle \zetawf \zetawf \zetawf \rangle
		\sim \langle \zetawf \zetawf \rangle^{3/2} .
	\end{equation}
	Assuming there are no other non-Gaussian contributions, $\fnl$ is given
	by the ratio 
	\begin{equation}
		\label{e:fnlest}
		\fnl = \frac{\langle \zetawf \zetawf \zetawf \rangle}
			{\Ps_{\rm total}(k)^2}
		\sim \frac{\Psw(k)^{3/2}}{\Ps_{\rm total}(k)^2} ,
	\end{equation}
	where $\Ps_{\rm total}$ is the total power spectrum of $\zeta$, 
	including the usual inflationary contributions as well as those from
	the waterfall process. The contributions from the waterfall need not
	dominate $\Ps_{\rm total}$.
	Fig.~\ref{fig1} shows
	that when $\sigma_\chi$ varies over $\sim 30$ orders of magnitude,
	$\sigmaw$ varies only over a factor of roughly two.  We will take
	a conservative point of view and estimate that the variance in $\zetawf$
	is given approximately by
	\begin{equation}
		\label{e:zzest}
		\Psw \sim \langle\zetawf\zetawf\rangle \sim \gamma
		\left( \frac{H}{\mu} \right)^2, \quad \mbox{with} \quad
		\gamma \sim 10^{-2} .
	\end{equation}
	The dependence on $H/\mu$ follows from our semi-analytic arguments and is
	supported by our numerical evidence.  The dimensionless factor 
	$\gamma \sim 10^{-2}$ arises in the same way.	From Fig.~\ref{fig1} 
	it is evident that	$\gamma$ varies from about
	$0.03$ -- $0.1$ over many decades of $\sigma_\chi$, so
	Eq.~\eref{e:zzest} is conservative.
	In practice, $\gamma$ is slightly larger in all
	of the explicit examples we have studied.
	In what follows, we
	quote requirements on $\fnl$ in terms of constraints on $H/\mu$.
	At the end of this section we will relate these constraints to parameters
	of the inflationary model such as $\lambda$, $V$, and $\epsilon$.
	
	In models where tachyonic preheating does not quench the
	production of $\zetawf$,
	the power spectrum and $\fnl$ put tight constraints on $H/\mu$.
	Whatever the other predictions of a given model of	 
	hybrid inflation, the power spectrum of $\zeta$ must match observation.
	Therefore its value is fixed at $\Ps_{\mathrm{total}} \sim 3 
	\times 10^{-9}$.
	If $\Psw$ makes
	a dominant contribution to $\Ps_{\mathrm{total}}$ the model is
	experimentally ruled out, because
	a correct normalization of the primordial power spectrum implies
	$\fnl \sim 	\Ps_{\mathrm{total}}^{-1/2} \sim 2 \times 10^4$.
	This is in grave
	conflict with experiment \cite{Smith:2009jr},
	which requires $|\fnl| \lesssim 100$. 
	To avoid $\zetawf$ dominating the power spectrum,
	the estimate of Eq.~\eref{e:zzest} shows that
	we must have $\Psw \lesssim \Ps_{\mathrm{total}}$,
	corresponding to $(H/\mu) \lesssim 5 \times 10^{-4}$.	
	
	In the regime where $\zetawf$ does not dominate $\Ps_{\mathrm{total}}$,
	a slightly stronger constraint on $H/\mu$ arises from
	considering $\fnl$.
	Combining Eqs.~\eref{e:fnlest} and~\eref{e:zzest}, and
	neglecting the contribution of $\Psw$ to $\Ps_{\mathrm{total}}$,
	we find
	\be
		\fnl \sim 10^{14} \left( \frac{H}{\mu} \right)^3
	\ee
	Since current experimental constraints require $\fnl \lesssim 10^2$, 
	we obtain the constraint 
	\be\label{e:fnlconstraint}
	\frac{H}{\mu} \lesssim 10^{-4}.
	\ee
	This is slightly stronger than the constraint from the power spectrum
	alone.	If future constraints on $\fnl$ improve to $|\fnl| \lesssim 1$,
	and we take a less conservative estimate that 
	$\Psw \sim 10^{-1} (H/\mu)^2$,
	the resulting constraint on $H/\mu$ would be an order of magnitude 
	tighter.
	
	\subsection{Constraints on hybrid model parameters}
	
	The constraints on $H/\mu$ turn out to be surprisingly restrictive when
	written in terms of the fundamental inflationary model parameters 
	$\lambda$, $V$, and $\epsilon$.	 Using the Friedmann equation, the
	slow-roll condition for $\phi$, and the definitions 
	\eref{potSimple} of $\lambda$ and \eref{eq:mu-def} of $\mu$,
	we find
	\be\label{e:HmuParams}
		 \frac{H}{\mu}	=
		\left(
		\frac{V_{\rm end}}{6 \sqrt{2} \Mp^3 \lambda \sqrt{\epsilon_{\rm end}}}
		\right)^{1/3}
	\ee
	Here, $V_{\rm end}$ and $\epsilon_{\rm end}$ 
	are the potential and slow-roll parameter evaluated
	at the end of  inflation,
	which corresponds to the beginning of the hybrid transition.	
	(Recall also that $\epsilon$ is defined using the derivative of $V$ in the
	$\varphi$ direction only, just before the hybrid transition).
	In the following,
	we will find it convenient to parametrize
	$V_{\rm end}$ as
	\be
		V_{\rm end} = M_I^4
	\ee
	where $M_I$ is the inflationary mass scale, measured at the end
	of inflation.
	Since current cosmological scales have exited the horizon long before
	the end of inflation, there are 
	no \emph{a priori} constraints on $M_I$ and $\epsilon_{\rm end}$
	from experiment. 
	
	We can get an idea of how strongly non-Gaussianity constrains
	the underlying hybrid inflationary model using a simple estimate, 
	in which we
	make certain assumptions about
	the naturalness of the potential.
	On CMB scales the 
	normalization of the scalar
	power spectrum constrains a combination of $M_I$ and $\epsilon$,
	specifically
	\be\label{e:Veconstraint}
		\frac{V}{24 \pi^2 \Mp^4 \epsilon} = \Ps_{\rm total} = 3 \times 10^{-9}.
	\ee
	Because $H$ and $\epsilon$ are close to constant during a hybrid
	phase
	it would be unnatural for this ratio to change significantly
	by the end of inflation,
	although such behaviour could be introduced by a 
	tuned or highly featured potential.
	Moreover, $\epsilon$ does not approach unity
	towards the end of inflation, because there is no graceful
	exit in a hybrid model. Therefore,
	we generically
	expect \eref{e:Veconstraint} to hold, roughly, throughout
	the hybrid phase.
	(For example, it holds for the potential
	\eref{e:numV} for reasonable choices of parameters.)
	Accordingly, we will take 
	$M_I \sim (3 \times 10^{-2}) \epsilon_{\rm end}^{1/4} \Mp$.
	This immediately implies that $\fnl \lesssim 100$ requires 
	$M_I^2 \lesssim \lambda 10^{-8} \Mp $.
	$M_I$ sets the scale of the inflationary potential, and 
	even if $\lambda$ took a value close to the Planck mass 
	this inequality would imply a rather low value for $M_I$.
	It is more reasonable to assume that $\lambda$ would take a value at a
	similar scale to $M_I$ --- otherwise we would be forced to introduce
	a hierarchy 
	between these mass parameters.	Taking $\lambda \sim M_I$ implies the
	constraint
	\begin{equation}
		M_I \lesssim 10^{-8} \Mp .
	\end{equation}
	The $\fnl$ parameter scales linearly with $M_I$, so if future experiments
	constrain $|\fnl| \lesssim 1$, this constraint tightens to 
	$M_I \lesssim 10^{-10} \Mp$.
	We conclude that
	unless the inflationary scale is very much lower
	than the Planck scale, unacceptable non-Gaussianity will be introduced.
	
	The constraints we have derived here should be viewed as guidelines
	for reasonable hybrid inflationary models.	If one has a particular
	example in mind,
	one should first compute the dimensionless factor
	$\gamma$
	appearing in \eref{e:zzest} --- although
	in practice we find this parameter varies very little between
	different potentials.
	One must then apply the definition of 
	$\fnl$ in \eref{e:fnlest} to find the corresponding constraint on
	$H/\mu$.  The formula \eref{e:HmuParams} constrains
	the underlying inflationary parameters $V$, $\lambda$, and
	$\epsilon$.	 The resuting constraints may be summarized by
	\be
	\Bigg{(} \frac{\fnl}{ 1.3 \times 10^4} \Bigg{)}\sim
	\Bigg{(}\frac{\gamma}{10^{-2}}\Bigg{)}^{3/2}
	\Bigg{(}\frac{M_I}{10^{-3} \Mp}\Bigg{)}^4
	\Bigg{(}\frac{10^{-2} \Mp}{\lambda}\Bigg{)}
	\Bigg{(}\frac{10^{-2}}{\epsilon_{\rm end}}\Bigg{)}^{1/2} ,
	\label{eq:total-constraint}
	\ee
	where we have assumed that $H/\mu$ is small enough that the waterfall
	makes a subdominant contribution to the $\zeta$ power spectrum.
	While there may be models for which Eq.~\eref{eq:total-constraint}
	is effectively unconstraining --- especially hybrid models with a very low
	inflationary scale and a correspondingly tiny 
	$\epsilon$ --- our analysis indicates that there are surprisingly
	strong constraints on a broad family of hybrid models.
	Moreover, these constraints can be expected to improve
	as future experimental results become available.
	
	\section{The validity of the separate universe picture}
	\label{sec:validity}

	In this section we return to the question posed at the
	beginning of \S\ref{sec:ps-and-fnl}, namely whether
	the trajectories followed by adjacent
	cosmological patches
	can be prevented from dispersing sufficiently to generate an
	appreciable $\zetawf$.
	Felder {\etal} \cite{Felder:2000hj} and later
	Felder, Kofman \& Linde \cite{Felder:2001kt}
	argued that a phenomenon known as
	\emph{tachyonic preheating} typically occurs in
	hybrid models.
	Lattice simulations of the effect show
	a rapid destabilization of the background evolution
	owing to gradient instabilities formed as the universe reheats on
	small scales.
	This destabilization invalidates the separate universe assumption.
	To determine how much dispersion occurs,
	we must decide whether CMB-scale regions of the
	universe reach kinetic domination before or after
	the onset of gradient instabilities.
	
	Which are the relevant timescales?
	To avoid overproduction of topological defects and black holes
	it is already known that the hybrid transition should complete
	in of order one e-fold \cite{GarciaBellido:1996qt}.
	This constraint is independent of $\zetawf$.
	Accordingly
	the shortest wavelength modes, which
	leave the cosmological horizon immediately prior to the tachyonic
	transition, must
	reach kinetic domination after $\lesssim 1$ e-fold.
	Their effective
	release point $\chi^{\rm (short)}_\ast$ 
	can be found using~\eref{varianceChi} and~\eref{asyWaterSol}
	with $k = k_\star$, 
	giving approximately
	\begin{equation}
		\chi^{\rm (short)}_\ast = \sqrt{\frac{C H^3}{6\pi^3 M}} .
		\label{eq:short-release}
	\end{equation}
	We
	recall that $C$ is implicitly defined in Eq.~\eref{eq:PB}, 
	and is of the form
	$C \sim a + b (M/\mu)^2$, where $a$ and $b$ are roughly unity.
	Together with~\eref{eq:transit-time},
	Eq.~\eref{eq:short-release} allows us to estimate
	the number of
	e-foldings, $N_S$, for the shortest wavelength modes to reach
	kinetic domination. It is
	\begin{equation}
		N_S
		\sim
		\frac{H}{\mu}
		\left(
			\ln \frac{6\pi^3 M}{HC}
		\right)^{2/3} .
		\label{eq:n-short}
	\end{equation}
	The analysis of Garc\'{\i}a--Bellido {\etal} \cite{GarciaBellido:1996qt}
	shows that the logarithm cannot be larger than $(H/\mu)^{-3/2}$.
	We also define $N_L$ to be the number of e-foldings required for
	CMB scales to reach kinetic domination.  These scales
	left the horizon $N_\ast$ e-foldings before the end of inflation,
	and hence the CMB scale is
	roughly $\e{N_\ast}$ larger than the horizon scale.
	In typical models
	$N_\ast$ is of order $50$ -- $75$.
	The scaling of $\chi_\ast$ with $k/k_\star$ indicates that
	\be
		\chi^{\rm (long)}_\ast = e^{-3 N_\ast / 2} \chi^{\rm (short)}_\ast.
		\label{eq:long-release}
	\ee
	Hence,
	\begin{equation}
		N_L
		\sim
		\frac{H}{\mu}
		\left( 3 N_\ast + 
			\ln \frac{6\pi^3 M}{HC}
		\right)^{2/3} .
	\end{equation}
	
	In typical models, the evolving field
	rolls into its reheating minimum almost immediately
	after kinetic domination.
	When it does so,
	Felder {\etal} \cite{Felder:2000hj,Felder:2001kt} argued that
	a rapid growth in occupation numbers of low-lying $\chi_{\vect{k}}$
	modes would efficiently drain energy from the rolling field, so that
	preheating would complete within a single oscillation.
	On this basis, we should expect the separate universe picture to
	provide an accurate description up to $\sim \mathrm{few} \times N_S$
	e-folds from the hybrid transition, but to fail at later times.
	A definitive determination of the range of
	validity of the separate universe picture will likely require a
	full lattice simulation of the waterfall phase.
	As a proxy for this unknown condition, we will assume that
	if $N_L/N_S \lesssim 10$ then
	there is a reasonable basis for belief that
	kinetic domination on CMB scales occurs sufficiently quickly
	for $\zetawf$ to be synthesized.

	There are two regimes in which the
	estimate of $N_L/N_S$
	is especially simple, depending on the relative size of the
	two parenthesized terms in \eref{eq:long-release}.
	Although one might have expected that the exponential suppresion
	of CMB-scale waterfall fluctuations would
	imply $N_L \gg N_S$, we will find that
	in both regimes,
	there is good evidence that the synthesis of $\zetawf$ can
	be successfully completed.
	The parity between $N_L$ and $N_S$
	improves when $H/\mu$ is made small enough to
	satisfy observational constraints.
	
	The first regime applies when
	$N_\ast \gtrsim \ln (6\pi^3 M/HC)$.  This regime is relevant when
	no large hierarchies exist between $M$, $H$, and $\mu$.
	In this case,
	\be
		N_L - N_S \sim \frac{H}{\mu} \left( 3 N_\ast \right)^{2/3},
		\hfill
		\mbox{for} \; N_\ast \gg \ln (6\pi^3 M/HC) .
		\hspace{1cm}
	\ee
	Both $N_L$ and
	$N_S$ are much smaller than unity, since they are proportional to
	the very small parameter $H/\mu$.
	In \S\ref{sec:ps-and-fnl}, we showed that consistency with 
	constraints on $\fnl$ requires $H/\mu \lesssim 10^{-4}$.  For 
	$N_\ast \sim 50-75$ this gives $N_L - N_S \lesssim 3 \times 10^{-3}$,
	showing that CMB-scale modes reach kinetic domination only
	a thousandth of an e-fold after short-scale modes.
	In this regime
	the ratio $N_L/N_S$ is controlled by the hierarchy between
	$N_\ast$ and the logarithmic factor, yielding
	\begin{equation}
		\frac{N_L}{N_S} \sim \left[ 1
			+ \frac{3 N_\ast}{\ln (6 \pi^3 M/HC)} \right]^{2/3}
		\hfill
		\mbox{for} \; N_\ast \gg \ln (6\pi^3 M/HC) .
		\hspace{1cm}
	\end{equation}
	Eq.~\eref{eq:n-short} shows that the logarithm may be of order $N_\ast$
	without spoiling the requirement $N_S \lesssim 1$.
	In the typical models studied in \S\ref{sec:evolution} we have found
	that $N_L/N_S$ is generally of order 5 -- 7. This
	analytic estimate is confirmed by our detailed Monte Carlo simulations.
	Pending detailed lattice calculations, there seems no reason to
	believe that perturbation theory should fail long before
	kinetic domination occurs on CMB scales.
	We also note that, while inflationary initial conditions imply
	that arguments based on causality give limited information,
	it may happen that the transition from an inflationary to
	a radiation-dominated equation of state takes some time to
	propagate from short to long scales. If so, it may be possible
	to tolerate larger values of $N_L/N_S$ than might seem
	reasonable on the basis of Refs.~\cite{Felder:2000hj,Felder:2001kt}.
	
	Even if a typical member of a bundle of CMB-scale trajectories does not
	reach kinetic domination, a small subset of trajectories in the bundle
	can be expected to do so. In this case the final curvature perturbation
	is likely to depend in detail on the tails of the distribution
	of $\chi_\ast$, and cannot be calculated in perturbation theory.
	We leave the exact details of this regime for future work.
	
	The second regime corresponds to $N_\ast \ll \ln (6\pi^3 M/HC)$, which 
	requires large hierarchies between $M$, $H$, and $\mu$.  In this
	regime, $N_L$ and $N_S$ are very nearly equal, with
	\be
		\label{e:nLnSregime2}
		N_L \sim N_S \sim \frac{H}{\mu} \left(
			\ln \frac{6\pi^3 M}{HC}
		\right)^{2/3}
		\hfill
		\mbox{for} \; N_\ast \ll \ln (6\pi^3 M/HC) .
		\hspace{1cm}
	\ee
	It follows that $N_S$ must be exceedingly small, since
	\be
		\frac{N_L - N_S}{N_S} \sim 
		\frac{2 N_\star}{ \ln (6\pi^3 M/HC) }
		\hfill
		\mbox{for} \; N_\ast \ll \ln (6\pi^3 M/HC) .
		\hspace{1cm}
	\ee
	Then, because
	$N_\star/\ln (6\pi^3 M/HC) \ll 1$ by assumption,
	the difference $N_L - N_S$ is itself much smaller than $N_S$.
	In this regime there seems a very good basis for belief that
	the separate universe assumption is not invalidated.
	Note that, however large the logarithm becomes,
	the requirement $N_S \lesssim 1$ can be satisfied by choosing
	a correspondingly small $H/\mu$.
		
	\section{Conclusions}
	\label{sec:conclusions}
	
	We have argued that the transition ending hybrid
	inflation can generate a large contribution to the curvature perturbation.
	This contribution arises from fluctuations in the 
	``waterfall'' field, $\chi$, which are created as modes exit the 
	inflationary horizon.
	These perturbations are subsequently suppressed by the
	cosmological expansion.
	Although these fluctuations are very small when inflation ends,
	they are amplified by the tachyonic instability in $\chi$
	and can provide a significant contribution to the curvature perturbation.
	To calculate the enhancement we use a gradient expansion in which
	we treat the moments of the initial distribution non-perturbatively.
	This method is similar to the method of ``moment transport'' which
	has recently been used to calculate the curvature perturbation in
	inflationary models \cite{Mulryne:2009kh}.
	Here, the procedure is simplified owing to our
	ability to calculate an approximate analytic solution.
	This technique predicts that
	the fluctuations generated by the waterfall
	are highly non-Gaussian, and can easily exceed observational
	constraints.
	If present in
	a given model, these effects lead to new constraints on viable
	hybrid models together with a unique observational signature.
	
	There are two phenomena which enable the tiny fluctuations 
	in $\chi$ to generate significant effects.
	First, the variance
	in $\delta N$ following completion of the waterfall
	depends extremely weakly
	on the variance in $\chi$ before the waterfall begins.  While
	surprising, this
	phenomenon is sufficiently general that it can be illustrated using an
	inverted simple harmonic oscillator.
	Similar behaviour is seen in the
	semi-analytic and numerical examples given above.
	Therefore,
	even very tiny fluctuations in $\chi$ can lead to large fluctuations in
	$\delta N$.  This means that the exponential suppression of
	$\chi$-fluctuations
	during inflation does not guarantee that these fluctuations
	are harmless when hybrid inflation completes.
	In a single-field model, 
	this kind of amplification would be forbidden because the curvature 
	perturbation $\zeta$ would be conserved on super-horizon scales.
	Since hybrid inflation necessarily involves multiple fields,
	the presence of
	isocurvature modes means that $\zeta$ is no longer conserved.
	As we have
	argued above, the hybrid waterfall can cause it to grow significantly.
	
	Second, the spectrum of fluctuations 
	in $\delta N$, and hence the spectrum of the curvature perturbation
	$\zeta$, is nearly scale-invariant.	Naively, one might
	have expected that any 
	process operating at the end of inflation could
	only affect modes of wavelength
	smaller than the horizon.
	This intuition is borne out for some processes:
	for example, the production of black holes or defects at the end of
	hybrid inflation results in a very blue power spectrum, which 
	would be unimportant on cosmological scales.  In
	the present case, it is true that $\chi$ has a 
	strongly blue power spectrum immediately prior to the hybrid transition.
	This is because fluctuations in $\chi$ are suppressed during 
	inflation. However, the waterfall process ensures that
	the final fluctution in $\delta N$ is nearly independent of that in
	$\chi$.  Therefore,
	the \emph{final} curvature fluctuation is
	nearly the same,
	independently of scale,
	even though it is seeded by fluctuations which,
	on long wavelengths, are much smaller.	
	Thus, the hybrid transition can reprocess the 
	blue spectrum of $\chi$ into a 
	nearly scale-invariant one.
	The key difference in behaviour between large and small scales is
	the transit time from the onset of the waterfall to kinetic domination.
	
	The analysis we have presented here depends crucially
	on the validity of the separate universe picture
	while CMB scales are being ejected from the ar\^{e}te.
	On small scales, it is known that the hybrid
	transition creates large (nonlinear) gradients in the 
	scalar fields. This rapid growth in gradients is responsible for the
	``tachyonic preheating" effect studied by
	Felder {\etal} \cite{Felder:2000hj,Felder:2001kt}.
	It is possible that 
	short wavelength (horizon-size) perturbations reach this nonlinear 
	regime very rapidly, before significant curvature perturbations are
	generated on longer wavelength (cosmological-size) scales.
	If so, this would prevent the dispersion of trajectories
	and lead to effects which are much
	weaker than those estimated here.
	An estimate of the onset of this gradient instability can be obtained
	by studying the time taken for different 
	scales to reach the kinetic-dominated regime, at which time the 
	curvature pertubation $\zeta$ ceases to evolve.
	Our estimates indicate that
	long and short wavelength perturbations reach the kinetic-dominated 
	regime at nearly the same time, within a tiny fraction of an e-folding
	from the onset of the hybrid transition.
	The hierarchy between the transit time for long and short
	wavelengths is typically a number of order unity.
	In this regime
	a subtle interplay exists between effects on short and long length
	scales, and the condition
	for validity of the separate universe picture is not known.
	We expect that a definitive resolution of these questions
	requires numerical simulations of the
	hybrid transition.
	We conclude that perturbation theory provides no compelling reason
	to believe that waterfall contributions to the curvature perturbation
	should be strongly suppressed.
	
	Even if tachyonic preheating prevents the generation of a significant
	curvature perturbation on CMB scales, it seems impossible to prevent
	the appearance of strong non-linearities on the shortest scales --- those
	associated with the horizon scale at the transition.
	If these fluctuations are large, they will collapse to form black
	holes or topological defects. The analysis we have presented
	indicates that the formation of these non-perturbative objects
	should be taken to proceed from very non-Gaussian initial conditions.
	It is known that the final mass fraction contained in such collapsed
	objects can depend on the detailed profile of initial fluctuations
	\cite{Bullock:1996at,Ivanov:1997ia,Hidalgo:2007vk},
	which potentially provides another means to constrain models
	containing a waterfall transition.
		
	\ack
	DS would like to thank the Center for Particle Cosmology at the 
	University of Pennsylvania for their hospitality during the completion
	of this work.	We would also like to thank	
	Christian Byrnes, Justin Khoury, Louis LeBlond, Jean-Luc Lehners,
	Paul Steinhardt, and Mark Trodden for useful comments and discussions.
	DS is supported by STFC. DM acknowledges support from the
	Centre for Theoretical Cosmology.
	
	\appendix

	\section{Evolution of a massive scalar field in de Sitter space}
	\label{appendix:massive}

	Here we give more details regarding the calculation of the power spectrum of
	$\chi$ before the tachyonic transition.	 This is used to determine 
	$\sigma_\chi^2(k)$, the variance of $\chi$ smoothed on a real space length
	scale of $2\pi/k$. 

	We assume that $\chi$ has the effective Lagrangian
	\be
		S = \int \left[ - \frac{1}{2} (\partial \chi)^2 + \frac{1}{2} M^2 \chi^2 
	\right]\; 
	\sqrt{-g}\,\d^4 x
	\ee
	and take $m$ to be constant.  We work in conformal time and assume that
	the spacetime is flat de Sitter space, and hence that
	\be
		\d s^2 = \frac{-\d \eta^2 + \d x_3^2}{(H \eta)^2}
	\ee
	where $H$ is the Hubble parameter.	As is conventional, we take $\eta$ to
	increase along $(-\infty,0)$ as proper time increases and the universe
	expands.
	We will find it convenient to use a
	rescaled field $u$, defined by
	\be
		u(t,\vec{x}) = (-H \eta) \chi(t,\vec{x})
	\ee
	Substituting this definition into the action for $\chi$, and decomposing
	$u$ into its Fourier modes, we find that each Fourier mode possesses the
	effective action 
	\be\label{ukaction}
		S = \int \frac{1}{2} (u_k')^2 - 
		\frac{1}{2} \left( k^2	- 
		\frac{2 - (M/H)^2}{\eta^2} \right) u_k^2 \; \d t
	\ee
	where $' = d/d\eta$.
	The rescaled field $u$ has a canonically normalized kinetic term in this
	action.	 This makes it simple to impose the usual Minkowski space
	boundary conditions well 
	inside the cosmological horizon.  The subtlety is that well outside the 
	horizon the mode functions are very different from their Minkowskian form.

	The equation of motion resulting from (\ref{ukaction}) is
	\be
		u'' + \left( k^2  - 
		\frac{2 - (M/H)^2}{\eta^2} \right) = 0
	\ee
	with solution
	\be
		u_k(\eta) = c_1 \sqrt{-k\eta} H^{(1)}_\nu(-k\eta) + 
		c_2 \sqrt{-k\eta} H^{(2)}_\nu(-k\eta)
	\ee
	where $H^{(1,2)}_\nu$ are Hankel functions and
	\be
		\nu = \sqrt{ \frac{9}{4} - \left(\frac{M}{H}\right)^2}
	\ee
	For applications to hybrid inflation, we require that the waterfall
	field is very massive, and in particular that $M \gg H$.  Hence the index
	of the Hankel function is purely imaginary.	 We assume that we work in this 
	limit throughout this derivation and hence take $\nu = i(M/H)$.
	We now canonically normalize the mode functions by taking  the 
	limit $-k \eta \to 0$ and using the small argument expansion\footnote{In 
	cosmological applications, usually one takes the
	``sub-horizon limit" of these 
	Hankel functions $H(x)$ to correspond to $x \gg 1$.	 In the present case of
	very massive fields with $M/H \gg 1$, the limit we require is actually that
	the proper field momentum is much larger than the inverse Compton 
	wavelength, which corresponds to $x \gg M/H = |\nu|$.  It is actually in 
	the limit $x \gg |\nu|$ that the Hankel functions can be approximated
	by an expression such as (\ref{HankelX}).}
	\be\label{HankelX}
		H^{(1,2)}_\nu (x) \to \sqrt{\frac{2}{\pi x}} \exp \left(
		\pm i x \mp \frac{i \pi \nu}{2} \mp \frac{\pi}{4} \right)
		\qquad x \gg |\nu| \gg 1
	\ee
	To canonically normalize the fields, we should have the positive frequency
	mode asymptote to $u_k(x) \to (2k)^{-1/2} e^{-ix}$ where $x = -k \eta$.	 
	Keeping in mind that
	$\mu$ is purely imaginary, this implies that the correctly normalized 
	positive frequency mode is described by
	\be
		c_1 = \frac{e^{-\pi |\nu| / 2}}{2} \sqrt{\frac{\pi}{k}}, \qquad c_2 = 0
	\ee
	To find the super-horizon limit, we take $x \ll 1$ and use
	\be
		H_\nu^{(1)}(x) = -\frac{i}{\pi} \left[
		\Gamma(-\nu) e^{- i \pi \nu} \left(\frac{x}{2}\right)^{\nu}
		+ \Gamma(\nu) \left(\frac{x}{2}\right)^{-\nu}
		\right] \qquad x \ll 1
	\ee
	Starting with the Euler reflection formula 
	\be
		\Gamma(-\nu) \Gamma(\nu) = - \frac{\pi}{\nu \sin (\pi \nu)}
	\ee
	and using the fact that $\Gamma(\nu)^* = \Gamma(\nu^*)$, the
	definition of $\chi$ in terms of $u$, and the assumption that
	$\nu \gg 1$, the super horizon
	limit of $\chi$ becomes
	\be
		\chi_k \to \sqrt{\frac{H^3}{k^3 M}} (-k \eta)^{(3/2)+ i(M/H) + i \theta}
		\qquad k\eta \gg 1, \quad M/H \gg 1.
	\ee
	where $\theta$ is an irrelevant constant phase angle.	Using this 
	expression,
	one can easily derive the correlators (\ref{ic2}) of $\chi_k$ on 
	super-horizon scales.

	\label{sec:hybrid}

	\section*{References}

\providecommand{\href}[2]{#2}\begingroup\raggedright\endgroup

\end{document}